\documentclass[%
 reprint,
 amsmath,amssymb,
 aps,
]{revtex4-2}

\usepackage{graphicx}% Include figure files
\usepackage{dcolumn}% Align table columns on decimal point
\usepackage{bm}% bold math
\begin{document}

\preprint{APS/123-QED}

\title{Efficient configuration-interaction models for photoionization of molecular dimers}% Force line breaks with \\


\author{Julio C Ruivo}
\author{Thomas Meltzer}%
\author{Alex G Harvey}%
\author{Jakub Benda}%
\author{Zdeněk Mašín}%
 \email{zdenek.masin@matfyz.cuni.cz}
\affiliation{%
Institute of Theoretical Physics, Faculty of Mathematics and Physics, Charles University, V Holešovičkách 2, 180 00 Praha 8, Czech Republic
}%

\date{\today}% It is always \today, today,

\begin{abstract}
We present R-matrix calculations of photoionization of molecular monomers and dimers, focusing on ammonia (NH$_3$) and formic acid (HCOOH), utilizing configuration-interaction models including the Occupation-Restricted Multiple Active Space (ORMAS) approach. We show that ORMAS is a highly efficient choice for calculating photoionization observables, yielding results that are in excellent agreement with those obtained using the much more demanding configuration-interaction method, Complete Active Space (CAS). We demonstrate that models incorporating single and double (SD) excitations with respect to the Hartree-Fock configuration provide good agreement with experimental data. The approach developed here can be readily applied to study photoionization in complex molecular systems.

\end{abstract}

\maketitle

%\tableofcontents

\section{Introduction}

The intricate interplay of photoinduced effects in nucleobases has long been a focal point in the study of DNA damage induced by ultraviolet (UV) light. Such reactions reveal various photochemical pathways, where electronic excitation serves as the primary event, and subsequent decay channels offer a spectrum of possibilities through nuclear dynamics and fragmentation.\cite{barbatti2015photoinduced} Among these pathways, the formation of long-lived excited states in nucleobase pairs emerges as a pivotal intermediate channel with reduced probability of radiative recombination. Consequently, they assume significance as potential actors in repair mechanisms, particularly in pyrimidine-type dimers.\cite{crespo2005base,crespo2008ground}
In addition to light-driven reactions, the interaction of low-energy electrons with DNA stands as a formidable agent in causing single and double strand breaks. This phenomenon is intricately linked to the formation of electron resonances localized on the nucleobases \cite{boudaiffa2000resonant}. The resonant states, inherent to all nucleobases, can undergo diverse decay processes, including elastic scattering, dissociative electron attachment leading to base release or fragmentation, and excitation of the base to an electronically excited state, which may also result in dissociation \cite{alizadeh2015biomolecular}. Low-lying resonances have been identified in all nucleobases \cite{winstead2008resonant, martin2004dna}, recent discoveries pointing to higher-lying resonances whose decay routes are not yet fully understood \cite{gorfinkiel2017electron}.

The alternative photoionization and electron impact processes in nucleobase pairs or nucleotides remain poorly understood theoretically due to the size of these molecules and the exponential scaling of the computational effort with system size. Thus, the complexities inherent in these systems necessitate the development of efficient configuration-interaction approaches to render calculations feasible, but without significant loss in the electronic correlation quality of description. Another challenge associated with such large systems is the absence of accurate experimental reference cross sections which makes benchmarking of theoretical calculations very difficult. One route, taken in this work, is to focus on molecular dimers and study the relationship between the properties of the dimer and the monomer, for which accurate reference data is available.

In this study, we provide calculated photoionization cross sections for ammonia (NH$_3$) and formic acid (HCOOH) in both monomer and dimer configurations. %s, and characterize resonance formation driven by the photoionization in these molecules. 
Due to its small size and number of electrons, ammonia is a straightforward candidate to carry out electronically robust photoionization calculations, and it can be benchmarked with experimental results \cite{cacelli1988photoionization}. Ammonia clusters, and especially dimers, can be used as a model for proton transfer.\cite{horke2016hydrogen, farmanara2002ultrafast} Formic acid monomer (FAM) has been used as a precursor of amino acids in the context of formation of $\pi^*$ resonances in electron collisions~\cite{janeckova2013,gallup2009}. Additionally, it has two sites supporting hydrogen bonds leading to a highly stable dimeric structure (FAD), therefore they are an ideal prototype to initiate studies of biomolecule pairs, including in photoionization.\cite{tenorio2019theoretical,tachikawa2020a} In a previous study,\cite{meltzer2022polarization} a total and state-resolved FAM photoionization cross section is presented in single- and multi-channel levels of description using the R-Matrix method, in good agreement with prior theoretical and experimental results, \cite{schwell2001he,fujimoto2020cross} emphasizing the importance of electron correlation in the photoionization description. To the best of our knowledge, there is no reported photoionization cross section data for FAD, except for the HeI photoelectron spectrum reported by Tomoda \textit{et al.}\cite{tomoda1983photoelectron} while fragmentation of FAD was studied more extensively by experiment.~\cite{heinbuch2007,hoshina2012} Therefore information on the primary photoionization event in FAD is lacking.%Multiple shape resonance formation in FAM, for at least 30 eV above the first ionization threshold, are shown in the calculated state-resolved cross sections, with width larger than 4 eV, but not so far identified.\cite{meltzer2022polarization} We characterize these resonances using scattering and Wigner time-delays.

Addressing the computational challenge inherent in calculating photoionization cross sections for dimers, we propose a model based on the occupation-restricted multiple active space approach (ORMAS) \cite{ivanic2003direct}, in which we specifically incorporate single and double (SD) excitations from a Hartree-Fock determinant as the reference. By concentrating on these essential electronic excitations, it is possible to substantially reduce the computational complexities associated with modeling continuum states in molecular dimers. To assess the effectiveness of our approach, we present a comparative analysis, contrasting results from ORMAS with those derived from a complete active space (CAS) level of calculation, the latter being applied to the ammonia structures. While a CAS model with 12 active electrons proved effective for FAM in prior studies,\cite{meltzer2022polarization} extending this approach to FAD is not computationally feasible. %Consequently, employing a more pragmatic SD model, incorporating twice the number of electrons (24 electrons) is largely opportune, with a computational feasibility while ensuring an acceptable level of accuracy in capturing adequate electron correlation.

Bridging this gap is vital not only for advancing our fundamental understanding of photoinduced processes in molecular dimers but also for developing a framework to investigate resonant states in broader contexts, such as nucleobase pairs. The text continues with Section 2, providing a concise overview of the R-Matrix method and the adopted models for photoionization calculations. In Section 3, we present our findings, including total and state-resolved photoionization cross sections, along with $\beta$-parameters. Section 4 is dedicated to a comprehensive discussion of these results. Finally, Section 5 encapsulates our conclusions, including outlooks for a continuation of this study.

\section{Theory and methodology}
Photoionization properties are obtained from the dipole transition matrix element,
\begin{equation}
    \langle \Psi_{f\mathbf{k_f}}^{(-)} \rvert \Vec{\mathbf{d}} \lvert  \Psi_i \rangle,
\end{equation}
where $\Psi_i$ and $\Psi_{f\mathbf{k_f}}^{(-)}$ are the respective initial and final wavefunction, and $\Vec{\mathbf{d}}$ is the dipole operator. In order to obtain the initial and final wavefunctions, we employed the R-matrix method to solve the Schrödinger equation for a fixed-nuclei molecule with $N$ electrons. 

\subsection{The R-matrix method}
The general formalism of the R-matrix method has been described in detail elsewhere~\cite{burke2011r,tennyson2010electron,mavsin2020ukrmol+} so we outline only a few relevant points here. The R-matrix method divides space into two regions, referred to as the inner and outer regions, defined by a boundary known as the R-matrix sphere. The two regions are solved separately using different computational methods, with the solutions matching at the boundary via the R-matrix. In the inner region, the Hamiltonian eigenstates are represented by a close-coupling expansion, 
\begin{equation}\label{eq:close-coupling_expansion}
    \begin{split}
        \psi_k^N  = \hat{A} &\sum_{i,j} c_{i,j,k}\Phi_i^{N-1}(\mathbf{r_1},...,\mathbf{r_{N-1}})\eta_{ij}(\mathbf{r_{N}}) \\
                    + &\sum_m b_{mk}\chi_m^{N}(\mathbf{r_1},...,\mathbf{r_{N}}).
    \end{split}
\end{equation}
In the equation above, the first term on the right-hand side represents the anti-symmetrized (by the $\hat{A}$ operator) product of the wave functions describing the target, $\Phi_i^{N-1}$, with continuum orbitals, $\eta_{ij}$. The second term  ($\chi^{N}$) comprises $L^2$ configurations where the ionized electron is placed in target orbitals, describing short range correlation/polarization. While the continuum functions are non-zero on the boundary, the $L^2$ configurations are confined to the inner region, and the charge density of bound states is zero in the outer region. Finally, the coefficients $c_{i,j,k}$ and $b_{mk}$ are obtained by diagonalizing the Hamiltonian in the inner region.\cite{mavsin2020ukrmol+} Typically, this is the most computationally demanding step, sometimes competing with molecular integral evaluation.

For all molecules, we shift the first ionization threshold to match the experimental value by applying a simple energy shift to the calculated cross section and beta parameter curves. The subsequent excited ionic states are shifted by the same amount to maintain the calculated relative energy positions.

\subsection{Smoothing procedure for inelastic results}
Photoionization observables computed from electronically inelastic models without further post-processing (called raw data in this work) typically include a high density of narrow spikes associated with autoionizing resonances or unphysical pseudoresonances~\cite{meltzer2022polarization,benda2020perturbative}. These spikes are usually unlikely to be observed in experiments due to experimental uncertainties, nuclear motion. Therefore this raises the challenge of generating smoothed curves suitable for comparison with experiment and to facilitate the comparison of the various models~\cite{benda2022analysis}. In this work smoothed curves were obtained by convoluting, separately, the absolute value $|d|$ and phase $\vartheta$ of the partial-wave dipoles in the polar form,
\begin{eqnarray}
    d_{flm}^{q}(E) = &&\int_{E_f}^\infty dE' g(E-E') |d_{flm}^{q}(E')|\nonumber \\&\times& e^{-i\int_{E_f}^\infty dE" G(E-E") \vartheta_{flm}^{q}(E")}
\label{eq:dipole_smooth}
\end{eqnarray}
where $f,l,m$ denotes the final cation state $f$ and its partial-wave quantum numbers, respectively; $E_f$ is the $f$-channel energy threshold; and $q$ is the component of the dipole transition. The convolution was performed with a Gaussian function $g(E-E')$ and the convoluted dipole transitions were normalized by the factor
\begin{equation}
    \int_{E_f}^\infty dE' g(E-E') = \int_{E_f}^\infty dE' \frac{1}{\sigma\sqrt{2\pi}}\exp\bigg[{-\frac{1}{2}\Big(\frac{E-E'}{\sigma(E)}\Big)^2}\bigg].
\end{equation}
The width $\sigma(E)$ is dependent on the photoelectron energy according to
\begin{equation}
    \sigma(E) = 3.10^{-5} + 0.03 \sqrt{\frac{E - E_{left}}{E_{right} - E_{left}}},
\end{equation}
where $E_{left}$ and $E_{right}$ specify the energy range for the smoothing and hartree units for energy are assumed in all formulas. In practice the upper limit in the energy integration coincides with the maximum energy requested in the calculation.

In previous R-matrix calculations the smoothing was performed separately on the real and imaginary parts of the dipole matrix element. While this approach gives reliable results in many molecules, it may lead to obviously distorted values of the magnitudes of the dipoles (and the cross sections) in certain cases. The polar form is more robust and ensures that the magnitude of the dipoles is preserved better. To show the effect of smoothing we always plot the raw data alongside the smoothed ones.

All calculations were carried out using R-matrix radius of 15 a.u. For the continuum description, a mixed Gaussian-type orbital (GTO) and B-spline-type orbital (BTO) basis was used with the maximum angular momentum $l_{max}$ determined by the convergence of the Static-Exchange photoionization cross section for each molecule. The GTO exponents were optimized for a radius 10 a.u., and a partially overlapping BTO basis was included starting from radius of 8 a.u. and extending up to the R-matrix radius. For each BTO partial wave we used a radial basis set consisting of 14 B-splines of order 6.

\subsection{Photoionization models}
We carried out the R-matrix calculations using the UKRmol+ package,\cite{mavsin2020ukrmol+} where the $N$-electron wavefunctions can be described using various models, including electron correlation on different levels. The simplest model to be used is the Static-Exchange (SE) model, a Hartree-Fock-based description that includes a single electronic channel of the ion and does not include electron correlation effects. SE model has been proven to qualitatively describe partial cross sections and angular distributions in cases where the HF approximation for the ion is valid, in particular for valence states. The polarization-consistent coupled Hartree-Fock (PC-CHF) model includes singly-excited HF-like ionic states and the corresponding set of polarization configurations.\cite{meltzer2022polarization} The PC-CHF model provides a more accurate description, however state-resolved properties tend to be inaccurate when configuration-interaction is required for the description of the target states and a more sophisticated treatment of electron correlation is needed. The complete active space model (CAS) consists in a more sophisticated description of electron correlation, where a set of active orbitals is selected to represent the electrons undergoing strong correlation effects, allowing all the possible excitations within the set. As would be expected, such a model is much more computationally demanding due to the combinatorial growth in the number of possible Slater determinants. 

In this work we extend the family of R-matrix scattering and photoionization models with the occupation-restricted multiple active space approach (ORMAS),\cite{ivanic2003direct} which uses the same definition of the active space as the CAS but includes only a subset of single and double (SD) and triple (SDT) excitations with respect to the ground HF configuration of the ion. With the ORMAS approach, the main electron correlation effects are preserved while configurations with secondary importance are dropped out. The ORMAS approach has been implemented into UKRmol-scripts~\cite{houfek2024} and allows an even finer control over the excitation subspaces and electrons distributed in them than explored in this work.

The basic challenge we face when extending the models from monomers to dimers is that of absence of reference cross sections for the dimers. To resolve this issue we apply the simple criterion that the total cross section for the dimer should have approximately twice the magnitude of the monomer cross section.

\section{Results}
In this section, we present and discuss the results obtained using the ORMAS strategy for calculating the photoionization cross sections and $\beta$-parameters of formic acid and ammonia in both monomer and dimer configurations. To assess the accuracy and efficiency of the ORMAS approach, we chose different levels of correlation (SD and SDT) and varied the number of active electrons. To benchmark our results, we performed calculations using the CAS model, where feasible.

%For all molecules, we shift the first ionization threshold to match the experimental value by applying a simple energy shift to the calculated cross section curves. The subsequent excited ionic states are shifted by the same amount to maintain the calculated relative energy positions. The smoothed curves were obtained by convolution of partial-wave dipoles with a Gaussian function, where the width of the Gaussian function is dependent on the photoelectron energy. 
%Raw curves reveal a high density of narrow spikes associated with resonances. While these spikes are unlikely to be observed in experiments, their presence raises the challenge of distinguishing whether they hold physical significance or are simply numerical artifacts.\cite{benda2020perturbative}

%All calculations were carried out using R-matrix radius of 15 a.u. For the continuum description, a mixed Gaussian type orbital (GTO) and B-spline type orbital (BTO) basis was used with the maximum angular momentum $l_{max}$ determined by the convergence of the SE photoionization cross section for each molecule. The GTO exponents were optimized for a radius 10 a.u. and a partially overlapping BTO basis was included starting from 8 a.u. to the R-matrix radius. For each BTO partial wave we used a radial basis set consisting of 14 B-splines of order 6.

\subsection{Photoionization of ammonia molecules}
Photoionization calculations of ammonia monomer were carried out for the equilibrium geometry of the neutral molecule, which belongs to the $C_{3v}$ point group symmetry but for practical reasons the calculations were performed in the $C_s$ point group symmetry. For the correspondent dimer, the calculations were performed for the $C_{2h}$ geometry, the second most stable conformation of the dimer structure, about 0.02 kcal/mol above the $C_s$ geometry.\cite{boese2003ab} This way, all-valence electron calculation could be performed for both molecules. The geometries were obtained with the MP2/aug-cc-pVDZ quantum chemistry method employing the adequate symmetry-adapted constraints. The calculations for ammonia monomer and dimer use HF orbitals optimized for the neutral ground state, using the Molpro package.\cite{werner2012molpro} The maximum angular momentum $l_{max}$ of 6 was employed in the continuum orbital expansion for both molecules.

Three models were employed for the monomer: CAS, SDT and SD, where the 8 valence electrons were distributed among 12 orbitals for both models, with 200 ionic states included in the first two models and the latter model included 132 ionic states (the maximum number allowed by the configuration space). The first four original calculated ionization thresholds $I_p$ are shown in Table \ref{tab:ip_ammonia_monomer} for each model, along with the dimensions of the ionic Hamiltonian matrices ($H^{N-1}$).% and the full $N$-electron Hamiltonian matrix ($H^{N}$) dimensions. 
The unshifted SD $I_p$ values deviate by approximately 3 eV from the experimental ones accompanied by a reduction of more than 99\% in the $H^{N-1}$ dimension compared to CAS. Aligning the first $I_p$ with the experimental value of 10.85 eV, as reported by Cacelli \textit{et al.}\cite{cacelli1988photoionization}, further reduces the deviation for the other states to approx. 1 eV from the experimental data. The inclusion of triple excitations in SDT brings $I_p$ to an excellent agreement with the CAS model, while still preserving a significant reduction of the number of configurations to 3\% of the CAS model.

%52\% in the $H^N$ dimension. Shifting the values for the three models further reduces the deviation to 1 eV from the experimental data. The inclusion of triple excitations in SDT brings $I_p$ to be in excellent agreement with the CAS model, but with a significant reduction to 3\% of the $H^{N-1}$ dimension and to 74\% of the $H^{N}$ dimension in the latter one.

\begin{table}[ht]
\centering
\caption{Ammonia monomer ionization thresholds (in eV) for the first four ionic states. Energies computed with the CAS, SD and SDT models using the (8,12) active space are compared with the experimental data of Cacelli \textit{et al.} \cite{cacelli1988photoionization} Energies obtained by shifting the first ionization threshold to the experimental result are shown in parentheses. The dimensions of the ionic Hamiltonian matrices are also shown in the bottom row and correspond to the sum of the Hamiltonian dimensions for both irreducible representations.}
\begin{tabular}{lllll} 
\hline
Orbital         & CAS(8,12)   & SD(8,12)   & SDT(8,12)  & Exp\cite{cacelli1988photoionization}  \\ \hline
4$a'$           & 12.1 (10.9) & 13.8 (10.9) & 12.0 (10.9) & 10.9 \\
3$a'$           & 18.1 (16.8) & 19.6 (16.6) & 17.7 (16.6) & 15.8 \\
1$a''$           & 18.1 (16.9) & 19.7 (16.8) & 17.9 (16.8) & --   \\
2$a'$           & 27.6 (26.4) & 30.3 (27.4) & 28.2 (27.0) & 27.7 \\ \hline
$H^{N-1}$ dim.  & 56,628      & 132         & 1,524       &      \\
%$H^{N}$ dim.    & 253,185     & 120,945     & 186,545     &      \\
\hline
\end{tabular}
\label{tab:ip_ammonia_monomer} 
\end{table}

For the dimer, we employed the models SD(16,12), SD(16,24) and CAS(12,10), both with 200 ionic states included. The first model corresponds to doubling the number of active electrons compared to the monomer but maintaining the same number of active orbitals, while the second model incorporates additional orbitals, allowing for more polarization and correlation to be accounted for. In contrast, the third model employs a smaller active space with fewer number of active electrons and orbitals, but with all possible electron occupations. For these results, the first ionization threshold was shifted to 10.35 eV in accordance with Farmanara \textit{et al.} \cite{farmanara2002ultrafast}
Table \ref{tab:ip_ammonia_dimer} presents the calculated $I_p$ for each model, along with the $H^{N-1}$ dimension. To our best knowledge, there is no experimental $I_p$ data available for excited ionic states of ammonia dimer, therefore the CAS results represent the best estimates we can provide. The unishifted thresholds from SD(16,12) are in a good agreement with CAS(12,10) but the model reduces the ionic Hamiltonian dimension by more than 99\%. The SD(16,24) model produced values about 2 eV above CAS with a reduction of the $H^{N-1}$ dimension by 96\%.

\begin{table}[ht]
\centering
\caption{Ammonia dimer ionization thresholds (in eV) for the first four ionic states. Energies were obtained with the CAS(12,10), SD(16,12) and SD(16,24) models. Energies obtained by shifting the first ionization threshold to the experimental result \cite{farmanara2002ultrafast} are shown in parentheses. The dimensions of the ionic Hamiltonian matrices are also shown in the bottom row and correspond to the sum of the Hamiltonian dimensions for all four irreducible representations.}
\begin{tabular}{llll} 
\hline
Orbital                  & CAS(12,10)  & SD(16,12)   & SD(16,24)   \\ \hline
4a$_\mathrm{g}$          & 12.0 (10.4) & 12.2 (10.4) & 14.1 (10.4) \\
4b$_\mathrm{u}$          & 12.1 (10.5) & 12.4 (10.5) & 14.2 (10.5) \\
3b$_\mathrm{u}$          & 17.0 (15.4) & 17.3 (15.4) & 19.3 (15.6) \\
1b$_\mathrm{g}$          & 17.7 (16.1) & 18.0 (16.1) & 19.9 (16.2) \\ \hline
$H^{N-1}$ dim.           & 27,720      & 264         & 1,032       \\
%$H^{N}$ dim.             & 226,270     & 216,275     & 222,383      \\
\hline
\end{tabular}
\label{tab:ip_ammonia_dimer}
\end{table}

\begin{figure}[ht]
    \centering
    \includegraphics[width=1.0\linewidth]{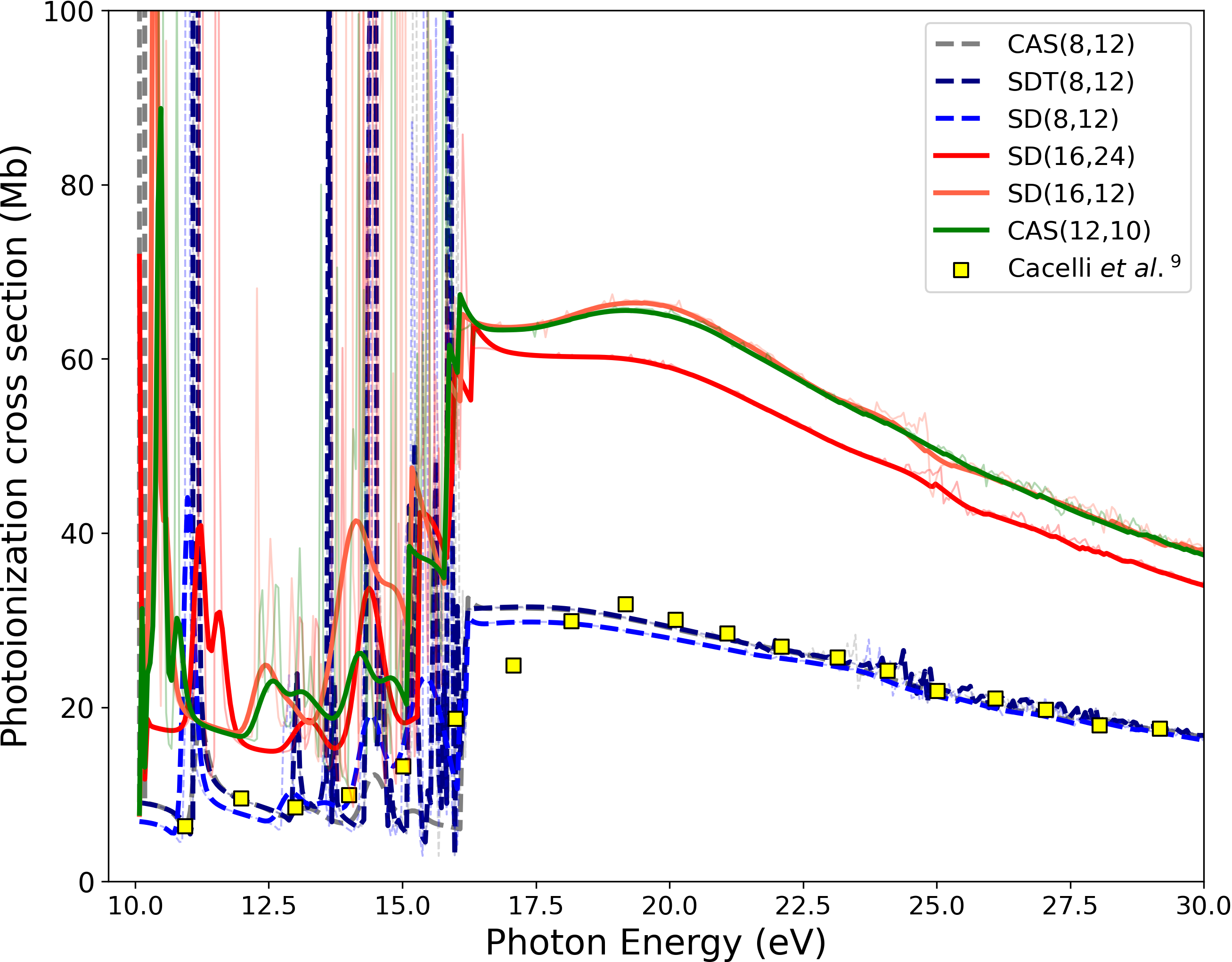}
    \caption{Total photoionization cross section of ammonia monomer and dimer. For the monomer, data from the CAS(8,12) (dashed gray), SDT(8,12) (dashed dark blue) and SD(8,12) (dashed blue) models are compared with the experimental results from Cacelli \textit{et al}\cite{cacelli1988photoionization} (yellow dots). Data from the SD(16,12) (solid orange), SD(16,24) (solid red) and CAS(12,10) (solid green) are shown for the dimer. Raw curves are shown with partially transparency.}
    \label{fig:tot_ixsec_ammonia}
\end{figure}

In Figure \ref{fig:tot_ixsec_ammonia}, we present the total photoionization cross sections for the ammonia monomer, calculated using both the CAS, SDT and SD models. For comparison, we include experimental data from Cacelli \textit{et al.}\cite{cacelli1988photoionization}
The calculated curves for the monomer exhibit a satisfactory agreement with experimental results. While the raw calculations reveal high-density spikes, the smoothed curves align more closely with the profile observed in the experimental data. Comparing the SD and CAS models, we observe an excellent agreement, though SD tends to be 5\% smaller than CAS above 16 eV. Notably, within the first 5 eV above the first ionization threshold, CAS produces a more pronounced background. Additionally, just above the second ionization threshold, SD and CAS demonstrate comparable magnitudes, but the SD curve is visibly reduced as the energy increases. The cross section obtained from SDT and CAS models show an excellent agreement, with marginal differences close to the 2nd $I_p$, indicating that SDT practically incorporates all the important correlation effects.

Figure \ref{fig:tot_ixsec_ammonia} includes the total cross section for the ammonia dimer calculated using the SD and CAS models. SD(16,12) demonstrates an excellent agreement with the CAS calculation, but variations are observed primarily in proximity to the ionization thresholds. On the other hand, SD(16,24) displays a magnitude approximately 10\% smaller than the other models, indicating that the inclusion of additional Hartree-Fock (HF) orbitals in the active space significantly changes the results. 

%Table \ref{tab:H_dim} shows the full $N$-electron Hamiltonian matrix dimension for the models in the monomer and dimer calculations. In the monomer, the SD model reduces in 52\% the $H^{N}$ dimension compared with the CAS model. The inclusion of triplet excitations results in the reduction of 26\%. In case of the dimer, the SD models allow to include more electrons and orbitals in the active space but keeping the order of magnitude in the Hamiltonian dimension.

Table \ref{tab:H_dim_nh3} presents the full $N$-electron Hamiltonian matrix dimensions for the models used in both monomer and dimer calculations. In the monomer, the SD model reduces the $H^{N}$ dimension by 52\% compared to the CAS model. The inclusion of triplet excitations achieves a reduction by 26\%. In the case of the dimer, the SD models allow for the inclusion of more electrons and orbitals in the active space while maintaining the same order of magnitude of $2\times10^5$ in the Hamiltonian dimension.

\begin{table}[ht]
    \caption{Full $N$-electron Hamiltonian matrix ($H^{N}$) dimension for the models employed in the ammonia monomer and dimer calculations.  The dimensions correspond to the sum of the Hamiltonian dimensions for all irreducible representations contributing to the bound-continuum dipole transition (two in the monomer and three in the dimer case, respectively).}
    \centering
    \begin{tabular}{llll}
    \hline
        \multicolumn{2}{c}{monomer} & \multicolumn{2}{c}{dimer} \\
        %monomer & model & dimer & model \\ \hline
        model & $H^{N}$ dim. & model & $H^{N}$ dim. \\ \hline
        CAS(8,12) & 253,385  & CAS(12,10) & 226,270  \\
        SD(8,12) & 121,077  & SD(16,12) & 216,275  \\
        SDT(8,12) & 186,745  & SD(16,24) & 222,383  \\ \hline
    \end{tabular}
\label{tab:H_dim_nh3}    
\end{table}

\begin{figure}[ht]
    \centering
    \includegraphics[width=1.0\linewidth]{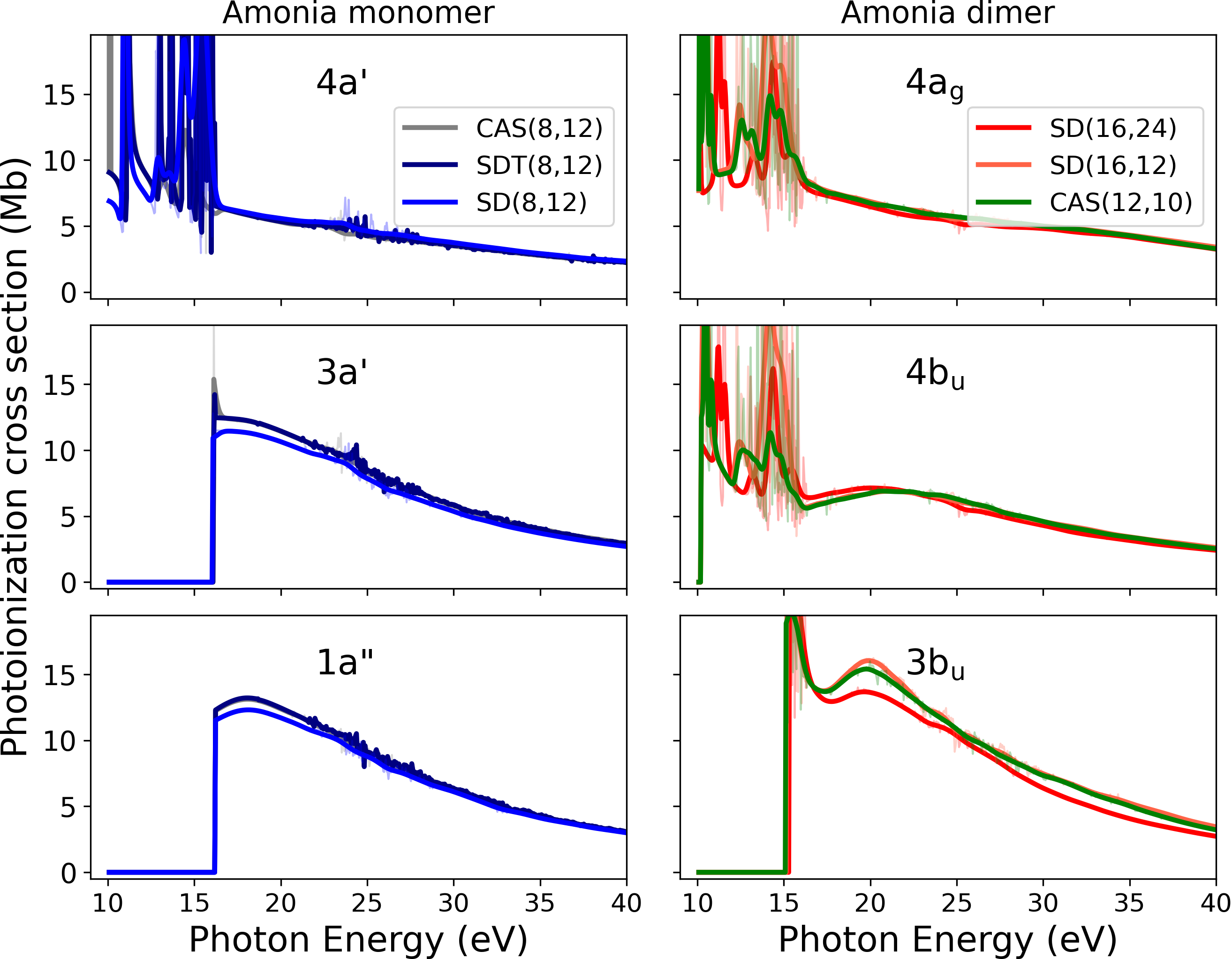}
    \includegraphics[width=1.0\linewidth]{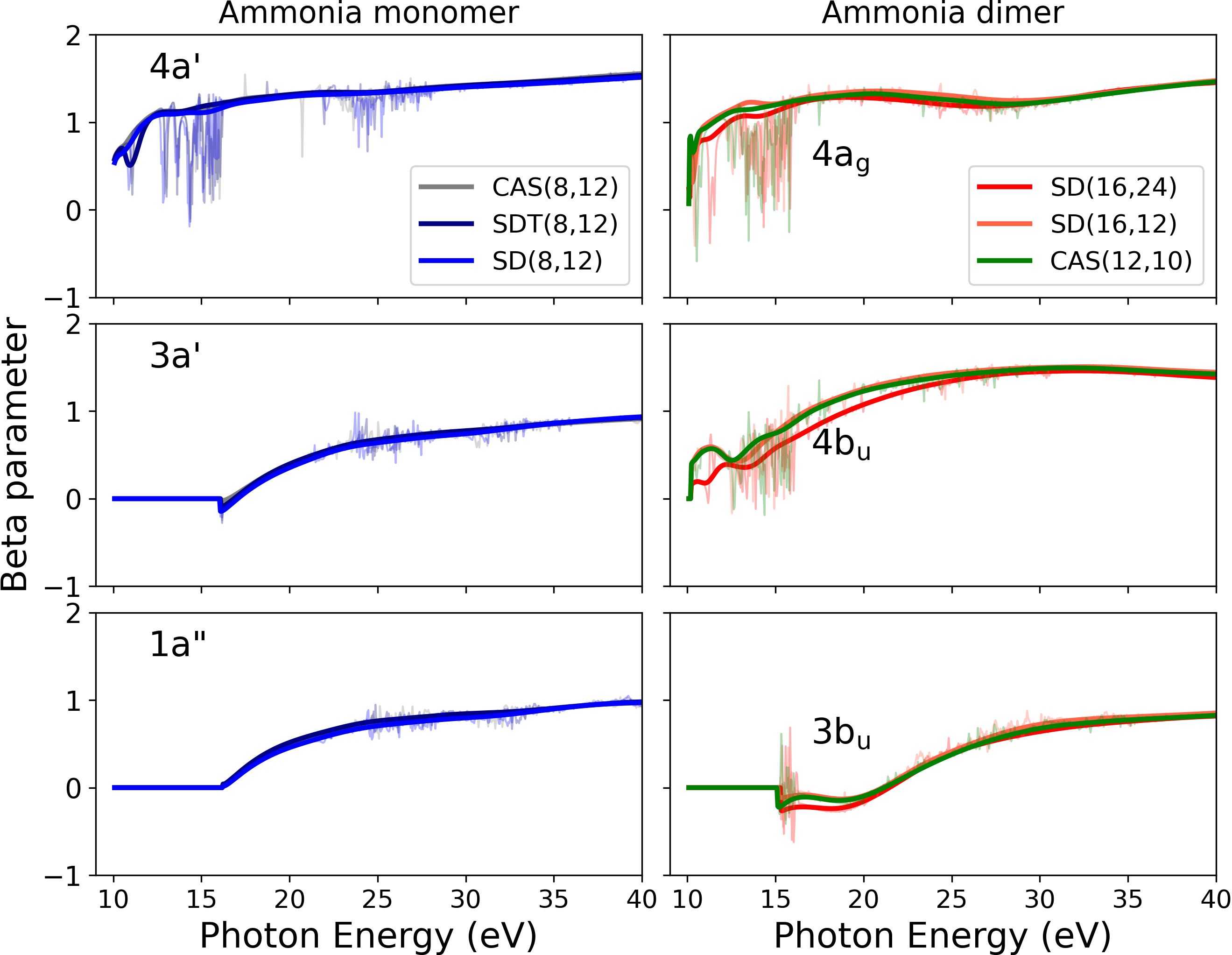}    
    \caption{State-resolved photoionization cross section (three panels in the top) and beta parameter (three panels in the bottom) of ammonia monomer (left) and dimer (right) for different models. States are labelled in according to ionized orbitals. Raw curves are shown with partially transparency.}
    \label{fig:parc_ixsec_ammonia}
\end{figure}

State-resolved photoionization cross sections and $\beta$-parameters for ammonia monomer and dimer are shown in Figure \ref{fig:parc_ixsec_ammonia}. The results show an excellent agreement between the CAS and SD models. In the case of the monomer, the cross section for the SD model is slightly smaller than that of CAS, a trend consistent with the observations in the total cross section. For the ammonia dimer, the SD(16,12) and CAS(12,10) models showcase an excellent agreement in both photoionization parameters, highlighting the consistency of outcomes between these models. However, the SD(16,24) model displays deviations from the other models, as evident in the total cross section.

%\begin{figure}[ht]
%    \centering
%    \includegraphics[width=1.0\linewidth]{images/beta_ammonia.png}
%    \caption{$\beta$-parameters of ammonia monomer (left) and dimer (right) for different models. States are labelled in according to ionized %orbitals.}
%    \label{fig:beta_ammonia}
%\end{figure}

%\clearpage

\subsection{Photoionization of formic acid molecules}
Photoionization calculations of FAM were carried out for the planar \textit{trans} conformer, the geometry of which ($C_s$ symmetry point group) was determined in our previous study.\cite{meltzer2022polarization} 
Target states were described using state-averaging (SA) CASSCF  orbitals. We adopted the same scheme employed in our previous study for the monomer, a SA-CASSCF(11,9) procedure with seven states included, the neutral ground state with a weight of 50\% and the first six ionic states with equal weight. The maximum angular momentum $l_{max}$ of 5 was employed in the continuum orbital expansion. 

We present results for FAM using three models: SD and SDT, in addition to the previous results\cite{meltzer2022polarization} obtained using the CAS model, all of them employing an active space of (12,9). In the CAS and SDT models, 200 ionic states were included, while 132 ionic states were included in the SD model. The first six calculated thresholds $I_p$ are shown in Table \ref{tab:ip_fam}, along with the $H^{N-1}$ dimensions. The first ionization threshold was shifted to the experimental value of 11.33 eV.\cite{leach2002photophysical}  The unshifted $I_p$ energies in the SD model deviate between 3 eV and 4 eV from the experimental values and the size of the ionic Hamiltonian is reduced by 98\%. The $I_p$ energies from the SDT model are in an excellent agreement with the CAS results to within 0.1 eV and the model is reduced by 87\% in the $H^{N-1}$ dimension.

\begin{table}[ht]
\centering
\caption{Formic acid monomer ionization thresholds (in eV) for the first six ionic states. Energies computed with the CAS,\cite{meltzer2022polarization} SD and SDT models using the (12,9) active space are compared with the experimental data.\cite{schwell2001he,leach2002photophysical} Energies obtained by shifting the first ionization threshold to the experimental result are shown in parentheses. The dimensions of the ionic Hamiltonian matrices are also shown in the bottom row and correspond to the sum of the Hamiltonian dimensions for both irreducible representations.}
\begin{tabular}{lllll}
\hline
Orbital      & CAS(12,9)\cite{meltzer2022polarization}         & SD(12,9)          & SDT(12,9)         & Exp\cite{schwell2001he,leach2002photophysical}   \\ \hline
10a$'$          & 12.5 (11.3) & 14.1 (11.3) & 12.6 (11.3) & 11.3  \\
2a$''$           & 13.6 (12.5) & 15.5 (12.7) & 13.6 (12.4) & 12.4  \\
9a$'$           & 16.3 (15.1) & 18.5 (15.7) & 16.3 (15.1) & 14.8  \\
1a$''$           & 17.2 (16.1) & 19.6 (16.8) & 17.3 (16.0)   & 15.4  \\
8a$'$           & 19.3 (18.1) & 21.3 (18.5) & 19.4 (18.2) & 17.0  \\
7a$'$           & 19.7 (18.5) & 22.2 (19.4) & 19.7 (18.5) & 17.3  \\ \hline
$H^{N-1}$       & 6,048       & 114         & 804         &     \\
%$H^{N}$ dim.     & 140,520     & 78,850      & 138,770     &     \\
\hline
\end{tabular}
\label{tab:ip_fam} 
\end{table}

For the calculations for the correspondent dimer, the equilibrium geometry ($C_{2h}$ point group symmetry) was taken from the theoretical results of Chocholouová \textit{ et al.}\cite{chocholouvsova2002potential}
The target orbitals were obtained using a (15,12) SA-CASSCF procedure with 13 states, the neutral ground state with the weight of 50\% and the first 12 ionic states with equal weight. The maximum angular momentum $l_{max}$ of 8 was employed in the continuum orbital expansion. 

We employed four models for the FAD calculations: CAS and SD, both with the (12,9) active space; and two SD models with the (18,15) and (24,18) active spaces, respectively. In the CAS model, as well as in the SD(18,15) and SD(24,18) models, we included 400 ionic states, while 114 ionic states were included in the SD(12,9) model. Table \ref{tab:ip_fad} includes the first eight calculated thresholds, along with the $H^{N-1}$ dimension. The experimental value for the first ionization threshold was 11.3 eV \cite{tomoda1983photoelectron}.
The $I_p$ energies in the CAS model are in the closest agreement with the experimental data, with deviations of approximately 2 eV. As the active space increases in the SD models, larger deviations are observed. However, the shifted $I_p$ values for these models are in a reasonable agreement with the experiment. Expanding the active space to include more electrons leads to better agreement in the higher-lying ionic states, primarily characterized by ionization from the orbitals not included in the CAS and SD(12,9) models. The $I_p$ energies for the SD(12,9) and CAS models are in an excellent agreement, while the SD model represents a reduction by 98\% in the $H^{N-1}$ dimension.

\begin{table}[t]
\caption{Formic acid dimer ionization thresholds (in eV) for the first eight ionic states. Computed energies are compared with the experimental data.\cite{tomoda1983photoelectron} Energies obtained by shifting the first ionization threshold to the experimental result are shown in parentheses. The dimensions of the ionic Hamiltonian matrices are also shown in the bottom row and correspond to the sum of the Hamiltonian dimensions for all four irreducible representations.}
%\begin{minipage}{\textwidth}
\centering
\begin{tabular}{llllll}
\hline
Orbital & CAS(12,9)   & SD(12,9)    & SD(18,15)   & SD(24,18)   & Exp\cite{tomoda1983photoelectron}  \\ \hline
10b$_\mathrm{u}$ & 13.1 (11.3) & 14.4 (11.3) & 16.5 (11.3) & 16.1 (11.3) & 11.3 \\
10a$_\mathrm{g}$ & 13.4 (11.6) & 14.8 (11.8) & 17.1 (12.0)   & 16.8 (12.1) & 12.0 \\
2a$_\mathrm{u}$  & 13.6 (11.9) & 14.9 (11.9) & 17.1 (11.9) & 16.7 (11.9) & 12.2 \\
2b$_\mathrm{g}$  & 13.7 (11.9) & 15.1 (12.1) & 17.4 (12.3) & 17.1 (12.3) & 12.7 \\
9b$_\mathrm{u}$  & 16.7 (15.0)   & 18.4 (15.4) & 20.9 (15.8) & 20.6 (15.8) & 14.0 \\
1b$_\mathrm{g}$  & 16.9 (15.2) & 18.7 (15.6) & 21.2 (16.0)   & 20.8 (16.1) & 15.2 \\
1a$_\mathrm{u}$  & 20.6 (18.8) & 22.3 (19.3) & 21.7 (16.5) & 21.7 (16.9) & 15.6 \\
9b$_\mathrm{u}$  & 17.9 (16.2) & 22.3 (19.3) & 24.3 (19.1) & 20.1 (15.3) & 15.6 \\ \hline
$H^{N-1}$  & 6,048       & 114         & 495         & 876         &     \\
%$H^{N}$ dim.      & 433,832     & 122,927     & 432,694     & 432,900     &   \\
\hline
\end{tabular}
\label{tab:ip_fad} 
%\end{minipage}
\end{table}

\begin{figure}[ht]
    \centering
    \includegraphics[width=1.0\linewidth]{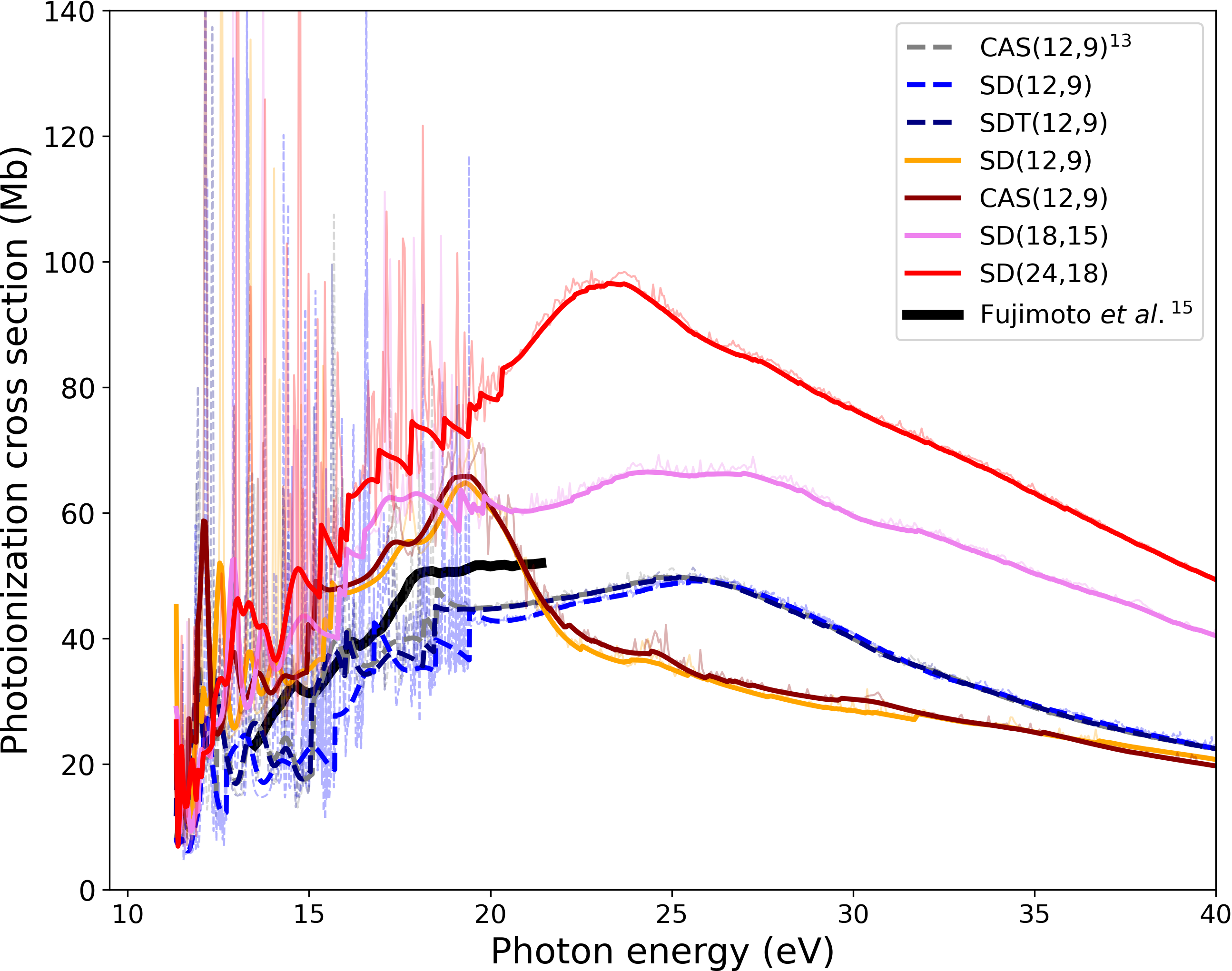}
    \caption{Total photoionization cross section of formic acid monomer (gray and blue and dark blue) and dimer (red, violet and yelow). Colours represent different models employed. Black dots indicate experimental results from Fujimoto \textit{et al.} (2020)\cite{fujimoto2020cross} Raw curves are shown with partially transparency.}
    \label{fig:tot_ixsec_formic_acid}
\end{figure}

In Figure \ref{fig:tot_ixsec_formic_acid}, we present the calculated total phoionization cross section for the formic acid monomer and dimer including the monomer experimental results of Fujimoto \textit{et al.}\cite{fujimoto2020cross} 
Our results, obtained using the SD(12,9) and SDT(12,9) models, are in excellent agreement with the previous CAS(12,9) calculation,\cite{meltzer2022polarization}
with minor deviations in the vicinity of the ionization thresholds. Additionally, our results also demonstrate good agreement with experimental data.\cite{fujimoto2020cross} 
Figure \ref{fig:tot_ixsec_formic_acid} also shows the calculated total photoionization cross section for the dimer structure. Our findings using the SD(12,9) and CAS(12,9) models are in excellent agreement, indicating the adequacy of the correlation included in the SD model. As more electrons and orbitals are included in the active space, the magnitude of the cross sections significantly increases, with the largest cross section coming from the SD(24,18) model. 

Table \ref{tab:H_dim_hcooh} presents the full $N$-electron Hamiltonian matrix dimensions ($H^N$) for the models used in our calculations. A CHF model employed in our previous calculation\cite{meltzer2022polarization} for FAM is also included in the comparison. In the monomer, the SD(12,9) model reduces the $H^N$ dimension to 56\% of the dimension of the CAS(12,9) model, while SDT(12,9) slightly reduces the $H^N$ dimension by 1.2\%. In comparison to the CHF model, CAS(12,9) corresponds to a 22\% smaller $H^N$ matrix dimension. In the dimer, SD(12,9) reduces the $H^N$ matrix dimension by 72\%. The SD(12,15) and SD(24,18) models generate the $H^N$ matrix dimension of the same order of magnitude $4\times10^5$ as the CAS(12,9) model. Nevertheless, the ionic calculation in the SD models is orders of magnitude faster and so are the individual diagonalizations of the $H^N$ matrix since the Table shows only the summed dimension over all irreducible representations.

\begin{table}[ht]
    \caption{Full $N$-electron Hamiltonian matrix ($H^{N}$) dimension for the models employed in the formic acid monomer and dimer calculations.  The dimensions correspond to the sum of the Hamiltonian dimensions for all irreducible representations contributing to the bound-continuum dipole transition (two in the monomer and three in the dimer case, respectively).}
    \centering
    \begin{tabular}{llll}
    \hline
        \multicolumn{2}{c}{monomer} & \multicolumn{2}{c}{dimer} \\
        %monomer & model & dimer & model \\ \hline
        model & $H^{N}$ dim. & model & $H^{N}$ dim. \\ \hline
        CAS(12,9)\cite{meltzer2022polarization} & 140,520  & CAS(12,9) & 433,832  \\
        SD(12,9) & 78,850  & SD(12,9) & 122,927  \\
        SDT(12,9) & 138,770  & SD(18,15) & 432,694  \\
        CHF a6v7\cite{meltzer2022polarization} & 179,740 & SD(24,18) & 432,900  \\ \hline
    \end{tabular}
\label{tab:H_dim_hcooh}    
\end{table}

\begin{figure}[ht]
    \centering
    \includegraphics[width=1.0\linewidth]{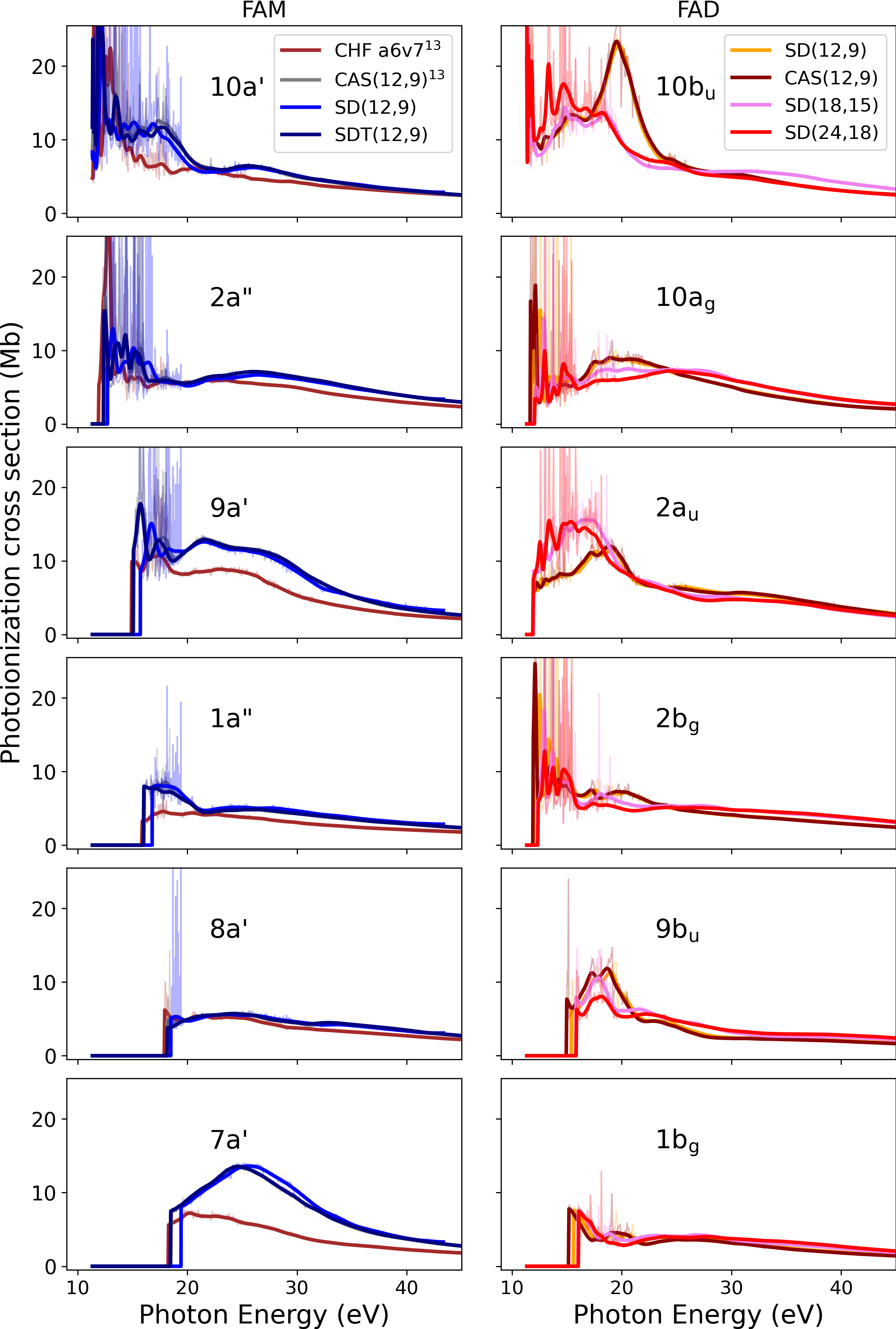}
    \caption{State-resolved photoionization cross section of formic acid monomer (left) and dimer (right) for different models. States are labelled in according to ionized orbitals. Raw curves are shown with partially transparency.}
    \label{fig:parc_ixsec_formic_acid}
\end{figure}

\begin{figure}[ht]
    \centering
    \includegraphics[width=1.0\linewidth]{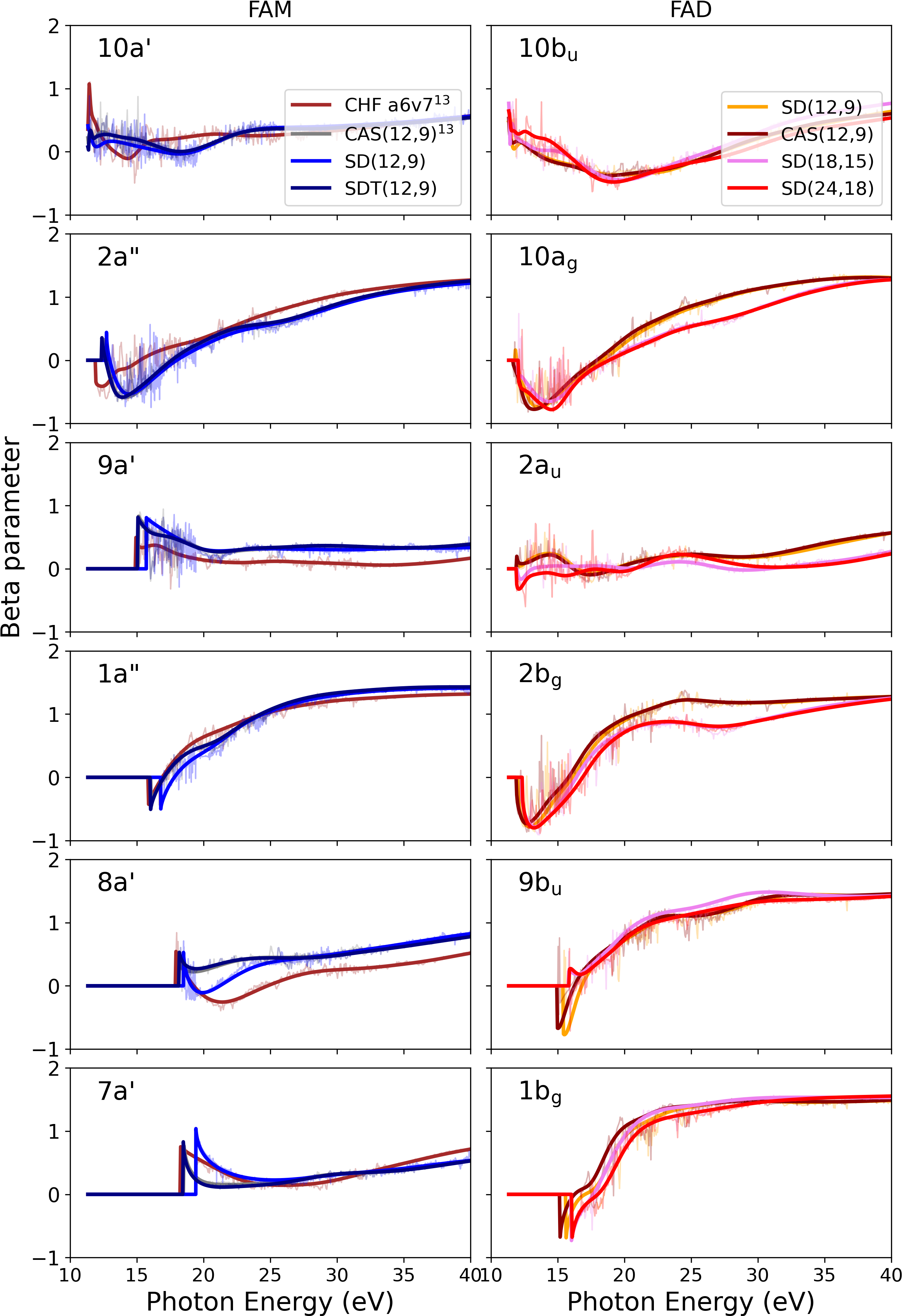}
    \caption{$\beta$-parameters of formic acid monomer (left) and dimer (right) for different models. States are labelled in according to ionized orbitals.}
    \label{fig:beta_formic_acid}
\end{figure}

State-resolved photoionization cross section for the formic acid monomer and dimer are shown in Figure \ref{fig:parc_ixsec_formic_acid}. The results show an excellent agreement between CAS(12,9), SDT(12,9) and SD(12,9) models in the monomer, as well as between CAS(12,9) and SD(12,9) in the dimer. The same observations apply to the state-resolved $\beta$-parameter calculations for FAM (Figure \ref{fig:beta_formic_acid}). As discussed by Meltzer and Mašín (2022),\cite{meltzer2022polarization} the CHF a6v7 model tends to perform worse for the higher excited ionic states where the correlation effects become important, but our results demonstrate that incorporating  double excitations is enough to adequately describe the electron correlation for this molecule. In the FAD calculations, CAS(12,9) and SD(12,9) shows larger magnitudes in the cross section close to 20 eV, while employing larger active spaces result in converged cross section in the whole energy range. Especially in 10b$_\mathrm{u}$ and in 9b$_\mathrm{u}$, structures are noticed in a more pronunciated shape for those models with the (12,9) active space. On the other hand, the cross section below 20 eV in 2a$_\mathrm{u}$ is larger for the SD(18,15) and SD(24,18) models, probably due to the higher density of autoionization resonance spikes described with these models. Interestingly, the $\beta$-parameters in FAD, presented in Figure \ref{fig:beta_formic_acid}, show only a relatively small variation between the models, except for the case of 10a$_\mathrm{g}$ and 2b$_\mathrm{g}$ states where we observe a non-negligible difference between the CAS and the SD (24,18) results in the region above 20~eV.

\section{Discussion}
Our results obtained using the ORMAS strategy have demonstrated an excellent agreement with those obtained using CAS models, both for ammonia and formic acid. This indicates that single and double excitations with respect to the ground HF configuration are sufficient to describe photoionization in these molecules. Meltzer and Mašín have pointed out that HF description of ionic states gives good results for low-lying states of formic acid, but such descriptions can be naturally expected to become inaccurate for the higher excited states.\cite{meltzer2022polarization} The use of the SD model recovers some of the correlation necessary for a good description of these states. The compatibility between CAS and SD in dimeric structures is also an excellent result, as doubling the system relative to the monomer would suggest the need to include quadruple excitations (two for each monomer, simultaneously). Thus, it is possible to include double the number of active electrons in calculations for dimers if the configuration space is restricted to double excitations.
The use of SD models facilitated a significant reduction in the dimensionality of the $N-1$ electron Hamiltonian, although at the expense of a systematic deviation of the ionization threshold values compared to the CAS or experimental data. However, the relative energies of the states are well preserved. Additionally, the SD models maintain the dimensionality of the full-electron Hamiltonian matrix, even with 24 electrons in the active space, at a level comparable to the corresponding CAS model with 12 electrons. As diagonalizing the Hamiltonian matrix is the most computationally demanding process, keeping its size under control is a crucial factor in the ability to produce results for larger polyatomic molecules and dimers or clusters.

Doubling the number of electrons included in the active space has proved crucial, as the magnitude of the total photoionization cross section was highly dependent on the number of active electrons. We have found that the total photoionization cross sections for the larger molecule, FAD, are more sensitive than ammonia dimer to the number of active orbitals included in the ORMAS model. Additionally, the simplest FAD models, based on CAS(12,9), give total photoionization cross sections which are significantly smaller than those generated by the larger ORMAS models, see Fig.~\ref{fig:tot_ixsec_formic_acid}. Applying the simple estimate that the cross section for the dimer can be expected to be approximately twice in magnitude, we rule the smallest models out as insufficient. It is the ORMAS models which double the number of active electrons that approximately satisfy this expectation.

Considering the results of the total photoionization cross sections for the dimers, their comparison with the results for the monomers reveals some interesting features, in particular in relation to resonance formation. Meltzer and Mašín suggested the resonance formation in FAM in at least one ionic state (10a$'$).\cite{meltzer2022polarization} Comparing our results for FAD and ammonia dimer, we note that several structures in different ionic states are associated with resonance formation, and comparison between them and the corresponding resonance formation in the monomers is imperative. This investigation will be presented in a subsequent work.

%The excellent agreement between the ORMAS-based models with CAS and experimental results proven the electon correlation in photoionization of both ammonia and formic acid is maily described by single and double electron excitations with respect to the ground HF configuration.  This assumption is extended to their dimeric structure,  and quadruple excitations (doubly promoted in each monomer) hardly percepted when the CAS and SD models are compared. On the other hand, our results are very sensitive to the active space size. The total cross section increasing in the FAD calculations when larger active spaces are employed demonstrates the importance to adequatly incorporing valence electrons. 

\section{Conclusions}
Our study presents the first theoretical investigation into the photoionization properties of ammonia and formic acid dimers, utilizing the occupation-restricted multiple active space (ORMAS) approach. We have demonstrated that ORMAS is a highly efficient strategy for calculating photoionization parameters, yielding results that are in excellent agreement with those obtained using the widely employed configuration interaction method, CAS. By employing models with a restricted active space, we were able to substantially reduce the dimensionality of the ionic states Hamiltonian while maintaining good agreement with the more expensive CAS results, particularly in maintaining the relative ionization threshold energies. We find that reliable dimer photoionization cross sections are generated in models which approximately double the number of active electrons compared to the monomer model. In case of formic acid dimer the ORMAS approach proved to be the only computationally feasible approach to generating reliable dimer cross sections.
%The photoionization calculation demands a significant computational effort due to the doubling of active electrons in the dimers and the increase in the number of configurations. Thereby, ORMAS provides an effective means to accurately describe photoionization processes in molecular dimers.
Moreover, our findings underscore the importance of considering electron correlation effects in accurately describing photoionization processes, particularly in higher excited states. We have shown that models incorporating single and double excitations with respect to the Hartree-Fock configuration provide good agreement with experimental data.
Our findings lay the groundwork for further investigations into resonance phenomena in molecular dimers and offer a new framework for exploring the photoionization properties of complex molecular systems.

%\section*{Author Contributions}
%We strongly encourage authors to include author contributions and recommend using \href{https://casrai.org/credit/}{CRediT} for standardised contribution descriptions. Please refer to our general \href{https://www.rsc.org/journals-books-databases/journal-authors-reviewers/author-responsibilities/}{author guidelines} for more information about authorship.

\section*{Conflicts of interest}
%In accordance with our policy on \href{https://www.rsc.org/journals-books-databases/journal-authors-reviewers/author-responsibilities/#code-of-conduct}{Conflicts of interest} please ensure that a conflicts of interest statement is included in your manuscript here.  Please note that this statement is required for all submitted manuscripts.  If no conflicts exist, please state that ``There are no conflicts to declare''.
There are no conflicts to declare.

\section*{Acknowledgements}
The authors acknowledge the support of the PRIMUS (20/SCI/003) project and the Czech Science Foundation (20-15548Y). Computational resources were supplied by the project “e-Infrastruktura CZ” (e-INFRA LM2018140) provided within the program Projects of Large Research, Development and Innovations Infrastructures. This work was also supported by the Ministry of Education, Youth and Sports of the Czech Republic through the e-INFRA CZ (ID:90254).
%%%END OF MAIN TEXT%%%


%%%REFERENCES%%%
\bibliography{apssamp} %You need to replace "rsc" on this line with the name of your .bib file

\end{document}